\documentclass[10pt,journal,twocolumn]{IEEEtran}
\usepackage{amssymb}
\usepackage{graphicx}
\usepackage{amsmath}
\usepackage{epsf}
\usepackage{mathrsfs}
\usepackage{cite}

\newtheorem{theorem}{Theorem}
\newtheorem{proposition}{Proposition}
\newtheorem{definition}{Definition}

\newtheorem{example}{Example}
\newtheorem{lemma}{Lemma}
\newtheorem{algorithm}{Algorithm}
\renewcommand{\QED}{\QEDopen}

\begin{document}

\title{To Decode the Interference or\\ To Consider it as Noise}
\author{ Abolfazl~S.~Motahari,~\IEEEmembership{Student Member,~IEEE,}
        and~Amir~K.~Khandani,~\IEEEmembership{Member,~IEEE}
\\\small Coding \& Signal Transmission Laboratory (www.cst.uwaterloo.ca)
\\ \{abolfazl,khandani\}@cst.uwaterloo.ca
}



\maketitle


\begin{abstract}
We address single-user data transmission over a channel where the received signal incurs interference from a finite number of users (interfering users) that use single codebooks for transmitting their own messages. The receiver, however, is allowed to decode interfering users' messages. This means the signal transmitted from any interfering user is either decoded or considered as noise at the receiver side.
We propose the following method to obtain an achievable rate for this channel. Assuming its own data is decoded successfully, the receiver partitions the set of interfering users into two disjoint subsets, namely the set of decodable users and the set of non-decodable users. Then the transmitter's rate is chosen such that the intended signal can be jointly decoded with the set of decodable users. To show the strength of this method, we prove that for the additive Gaussian channel with Gaussian interfering users, the Gaussian distribution is optimal and the achievable rate is the capacity of this channel.
To obtain the maximum achievable rate, one needs to find the maximum decodable subset of interfering users. Due to the large number of possible choices, having efficient algorithms that find the set of decodable users with maximum cardinality is desired. To this end, we propose an algorithm that enables the receiver to accomplish this task in polynomial time.
\end{abstract}


\section{Introduction}
\PARstart{T}{reating} interference as noise is not always the best strategy. Since the interference caused by another user is intrinsically a codeword from a codebook, it is possible to decode the interference at the receiver side which results in transmitting at higher rates. Whereas, if the interference is considered as noise, the structure of interference is ignored resulting in lower data rates.  Moreover, there are multi-user channels whose capacity regions are achieved when the interference is decoded by some or by all receivers. Examples of such channels are the degraded Broadcast Channel (BC) and strong two-user Gaussian Interference Channel (IC), c.f. \cite{cover:BC,STRONG:SATO,Very_Storng:Carleial}.

Decoding interference caused by interfering users is addressed in different papers.
As an example, the best inner bound for the two-user IC is the Han-Kobayashi achievable rate region \cite{IC:HK} which is obtained by combining the multilevel coding and joint typical decoding. Transmitters split their data into two parts where independent codebooks are used to transmit each part. Receivers jointly decode their own data and some part of the data by the unintended user.

Focusing on a particular transmit-receive pair in a multi-user system, we obtain a single-user channel where the received signal incurs interference from a number of interfering users. We assume that interfering users use single codebooks, which are generated independent of each other. Having information about the rates and codebooks of interfering users, the receiver is allowed to decode interfering messages. This in turn means that the signal transmitted from any interfering user is either decoded or considered as noise.


We propose the following method to obtain an achievable rate for the channel. Assuming its own data is decoded successfully, the receiver finds the maximum decodable subset of interfering users. By a maximum decodable subset, we mean a set of users that are decodable at the receiver, regarding the rest as noise and any decodable set is a proper subset of it. It is shown that this task can be accomplished by using a polynomial time algorithm. Once the receiver obtains the maximum decodable subset, it can partition the interfering users into two disjoint subsets, namely \emph{decodable users} and \emph{non-decodable users}. Then, the transmitter's rate is chosen such that the intended signal can be jointly decoded with the set of decodable users. We also propose a polynomial time algorithm to find the maximum achievable rate obtainable by this method.

To show the strength of this method, we prove that for the additive Gaussian channel with Gaussian interfering users, the Gaussian distribution is optimal and the achievable rate is the capacity of this channel.

As an application, we use this model to characterize some achievable rate points for the $M$-user Gaussian IC. In general, we are interested in characterizing achievable rate region for the $M$-user IC where each transmitter is allowed to transmit data by using a single codebook and each receiver is allowed to decode any subset of interfering users. Therefore, at each receiver, the signal transmitted from any interfering user is either decoded or considered as noise.

Despite the two-user case, the general $M$-user IC is less studied in the literature. The state of the art work for deriving achievable rate vectors treats interfering users as noise \cite{Gupta00,Etkin-Spectrum,Weber06,masoud,Gesbert07,Leveque05,Imhof05,Boche_convexity,Boche_iterative}.  For example, in \cite{Etkin-Spectrum} the $M$-user Gaussian IC is studied where transmitters are allowed to allocate different powers in different bandwidths and receivers treat interference as noise. Recently, in \cite{Mohammad-k-user,Steve-Weber}, successive interference cancelation is studied. For example, in \cite{Mohammad-k-user} the optimal order of decoding that maximizes the minimum rate among all users is obtained.

The organization of this paper is as follows. In Section II, we introduce the system model and some background materials. In Section III, we consider a discrete memoryless channel consisting of $M$ transmitters and one receiver. We assume that the users' rate vector is not necessarily inside the capacity region of the Multiple Access Channel (MAC) seen at the receiver side which results in failure of the receiver to reliably decode all the data streams. The receiver's task, however, is to find a maximum decodable subset of transmitters so that their data can be decoded from the received signal. We propose a polynomial-time algorithm which finds the maximum decodable subset of users.

In Section IV, we consider single-user data transmission over a channel with $M-1$ interfering users. We first obtain a lower bound and an upper bound on the capacity of this channel. Then, we propose a method that characterizes an achievable rate for the channel. This achievable rate is a function of other users' rates. We then prove that this function is piecewise linear.

In Section V, we consider additive channels where the interference caused by other users is Gaussian. We prove that for this case, the Gaussian codebook achieves the capacity where each interfering user is either decoded or treated as noise by the receiver.

In Section VI, we investigate applications of the proposed algorithms to the $M$-user Gaussian IC. We first develop a polynomial time algorithm that characterizes points obtainable from successive maximization of users' rates. We then generalize the notion of one-sided Gaussian ICs to the $M$-user case, and characterize a point on the boundary of the capacity region. We finally obtain the capacity of the strong one-sided $M$-user Gaussian IC. In Section VII, we conclude the paper.

\emph{Notations}: Throughout this paper, we use the following notations. Vectors are represented by bold faced letters. Random variables and sets are denoted by capital letters where the difference is clear from the context. The difference, union, and intersection of two sets $U$ and $V$ are represented  by $U\backslash V$, $U\cup V$, and $U\cap V$, respectively. The complement of a subset $U$ is denoted by $\overline{U}$. The cardinality of a set $U$ is denoted by $|U|$. $M$ is always the number of users in the system. We use $E$ to denote the set $\{1,2,\ldots,M\}$. $2^E$ denotes the power set of $E$ which is the collection of all subsets of $E$. For any set $S\subseteq E$ and any vector $\mathbf{x}=[x_1,x_2,\ldots,x_M]$, we use the compact notations $\mathbf{x}(S)$ and $\mathbf{x}_S$ to denote $\sum_{i\in S}x_i$ and $[x_i]_{i\in S}$, respectively. In particular, $\mathbf{x}_{-i}=\mathbf{x}_{\overline{\{i\}}}=[x_1,\ldots,x_{i-1},x_{i+1},\ldots,x_M]$. $\Re$ is the set of real numbers and $\Re^k$ denotes a $k$-dimensional Euclidean space.

\section{Preliminaries}

\subsection{System Model}
\begin{figure}
\centering \includegraphics[scale=0.75]{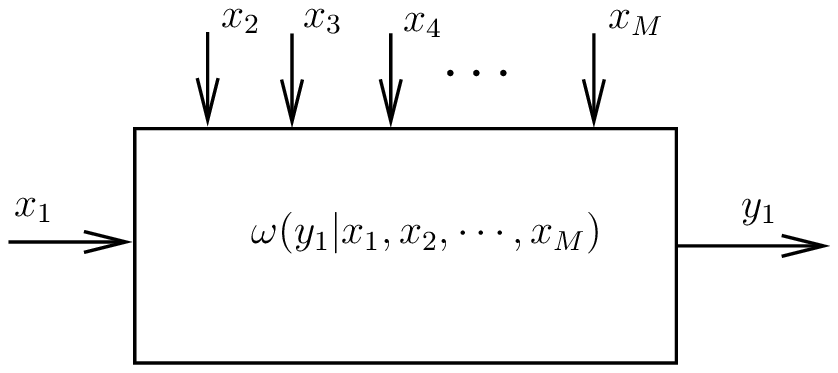}
\caption{Single user in an interfering medium. $x_1$ is the input letter from the intended user and $x_i$, $i=2,3,\ldots,M$, is the input symbol corresponding to the $i$'th interfering user.}\label{Fig Single}
\end{figure}

We consider single-user data transmission over a channel $\mathscr{S}$ with $M-1$ interfering users. $\mathscr{S}$ is specified by the transition probability function $\omega(y_1|x_1,x_2,\ldots,x_M)$ where $x_i\in\mathscr{X}_i$ is the input letter to the channel from the $i$'th user and $y_1\in\mathscr{Y}_1$ is the output letter received by the receiver, see Figure \ref{Fig Single}. The set of users' indices is denoted by $E$. $x_1$ is the input letter from the intended user and $x_i$ for $i=2,3,\ldots,M$ are input letters from interfering users. We assume that the interfering users transmit data at the rate vector $\mathbf{R}_{-1}=[R_2,R_3,\ldots,R_M]$ by using single codebooks generated randomly from the joint probability distribution $p_{X_2}(x_2)p_{X_3}(x_3)\cdots p_{X_M}(x_M)$. We are interested in characterizing the capacity of this channel.

We also consider the continuous Gaussian case modeled by
\begin{equation}\label{single-receiver}
y_1=x_1+x_2+\cdot\cdot\cdot+x_{M}+z,
\end{equation}
where $x_1$ and $y_1$ denote transmitted and received symbols, respectively. $x_i$, $i=2,3,\ldots,M$, is the input symbol corresponding to the $i$'th interfering user that uses a single Gaussian codebook with power $P_i$ and rate $R_i$. $z$ is the additive white Gaussian noise with variance $N$. The transmitter is subject to the average power $P_1$ and tries to send data at the maximum rate $R_1$.

\subsection{Submodular Functions}
\begin{definition}
Let $E$ be a finite nonempty set. A function  $f:2^E\rightarrow \mathbb{R}$ is called a submodular function if it satisfies
\begin{equation}
f(V\cup U)+f(V\cap U)\leq f(V)+f(U),
\end{equation}
for any $V,U\subseteq E$. A function $f$ is called suppermodular if $-f$ is submodular. A modular function is a function which is both submodular and suppermodular.
\end{definition}

Submodular functions are one of the most important objects in discrete optimization. In fact, they play the same role in discrete optimization as convex functions do in the continuous case \cite{Schrijver}. Besides having a polynomial-time algorithm based on the ellipsoid method \cite{lovasz-geometric}, there are combinatorial algorithms for minimizing submodular functions in strongly polynomial time, c.f. \cite{Schrijver} and \cite{IWATA}.

If a submodular function is nondecreasing, i.e. $f(U)\leq f(V)$ if $U\subseteq V$, and $f(\varnothing)=0$, then the associated polyhedron
\begin{equation}\label{polymatroid}
\mathcal{B}(f)=\{\mathbf{x}|\mathbf{x}(U)\leq f(U),~\forall U\subseteq E, \mathbf{x}\geq 0\},
\end{equation}
is a polymatroid. Likewise, if a suppermodular function is nondecreasing and $f(\varnothing)=0$, then the associated polyhedron
\begin{equation}
\mathcal{G}(f)=\{\mathbf{x}|\mathbf{x}(U)\geq f(U),~\forall U\subseteq E\},
\end{equation}
is a contra-polymatroid.

\subsection{Properties of Mutual Information for Independent Random Variables}
In this subsection, we review some important equalities and inequalities in Information Theory. We consider $M$ independent random variables $X_1,X_2,\ldots,X_M$. Moreover, let $E=\{1,2,\ldots,M\}$ denote the set of random variables' indices. For any random variable $Y$, we have the following properties:

1) \emph{Chain Rule}: For any disjoint subsets $U$ and $V$, we have the following inequality:
\begin{equation}
I(\mathbf{X}_{U\cup V};Y)=I(\mathbf{X}_V;Y|\mathbf{X}_{U})+I(\mathbf{X}_U;Y).
\end{equation}

2) \emph{Independent Conditioning Inequality}: For any disjoint subsets $U$ and $V$, the following inequality holds:
\begin{equation}\label{main-inequality}
I(\mathbf{X}_{U};Y)\leq I(\mathbf{X}_U;Y|\mathbf{X}_{V}).
\end{equation}

3) \emph{Polymatroidal Property}: In \cite{Han:correlated-mac}, it is shown that the set function $\sigma(U)=I(\mathbf{X}_U;Y|\mathbf{X}_{\overline{U}})$ is submodular and nondecreasing, i.e.,
\begin{equation}
\sigma(U\cup V)+\sigma(U\cap V)\leq \sigma(U)+\sigma(V),~\forall U,V\subseteq E.
\end{equation}
Hence, its associated polyhedron is a polymatroid.

4) \emph{Contra-polymatroidal Property}: We claim that the set function $\rho$ defined as $\rho(U)=I(\mathbf{X}_U;Y)$ is a suppermodular function. To this end, fix any arbitrarily subsets $U$ and $V$. Let $S=U\cap V$. From the chain rule, we have
\begin{equation}
I(\mathbf{X}_{U\cup V};Y)=I(\mathbf{X}_{U};Y)+I(\mathbf{X}_{V\backslash U};Y|\mathbf{X}_U),
\end{equation}
which can equivalently be written as
\begin{equation}
\rho(U\cup V)=\rho(U)+I(\mathbf{X}_{V\backslash S};Y|\mathbf{X}_{U\backslash S},\mathbf{X}_S).
\end{equation}
From Independent Conditioning Property, we have $I(\mathbf{X}_{V\backslash S};Y|\mathbf{X}_S)\leq I(\mathbf{X}_{V\backslash S};Y|\mathbf{X}_{U\backslash S},\mathbf{X}_S)$. Hence,
\begin{equation}
\rho(U\cup V)\geq \rho(U)+I(\mathbf{X}_{V\backslash S};Y|\mathbf{X}_S).
\end{equation}
Adding $\rho(U\cap V)=\rho(S)$ to both sides, we obtain
\begin{equation}
\rho(U\cup V)+\rho(U\cap V)\geq \rho(U)+I(\mathbf{X}_{V\backslash S};Y|\mathbf{X}_S)+I(\mathbf{X}_S;Y).
\end{equation}
Since $I(\mathbf{X}_{V\backslash S};Y|\mathbf{X}_S)+I(\mathbf{X}_S;Y)=I(\mathbf{X}_{V};Y)$, we have
\begin{equation}
\rho(U\cup V)+\rho(U\cap V)\geq \rho(U)+\rho(V),
\end{equation}
as claimed. It is easy to show that $\rho$ is nondecreasing and hence its associated polyhedron is a contra-polymatroid.

\subsection{Multiple Access Capacity Region}
One of the most important results in Information Theory is the characterization of the capacity region of the MAC \cite{Liao:MAC,Ahlswede:two-sender-two-receiver}. The capacity region of a MAC can be represented as follows. We define $\mathscr{P}$ as the collection of all probability distributions which can be written as $\mathbb{P}(x_1,x_2,\ldots,x_M,y)=p(x_1)p(x_2)\cdots p(x_M)\omega (y|x_1,x_2,\ldots,x_M)$, where $\omega (y|x_1,x_2,\ldots,x_M)$ is the channel transition probability function. Now, the capacity region of a MAC is
\begin{equation}
\mathcal{C}_{\text{MAC}}=\text{conv}\left(\bigcup_{\mathbb{P}\in \mathscr{P}}\mathcal{C}_{\text{MAC}}(\mathbb{P})\right),
\end{equation}
where $\text{conv}(\cdot)$ denotes convex hull operation, and $\mathcal{C}_{\text{MAC}}(\mathbb{P})$ is defined as
\begin{equation}\label{mac capacity}
\mathcal{C}_{\text{MAC}}(\mathbb{P})=\{\mathbf{R}|\mathbf{R}(U)\leq I(\mathbf{X}_U;Y|\mathbf{X}_{\overline{U}}),~\forall~U\subseteq E\}.
\end{equation}

Using the polymatroidal property of the mutual information, it is easy to show that $\mathcal{C}_{\text{MAC}}(\mathbb{P})$ is a polymatroid. It is worth noting that even though $\mathcal{C}_{\text{MAC}}$ is the union of polymatroids, it is not necessarily a polymatroid. However, $\mathcal{C}_{\text{MAC}}$ is a polymatroid for the $M$-user Gaussian MAC modeled by
\begin{equation}
y=x_1+x_2+\cdot\cdot\cdot+x_{M}+z,
\end{equation}
where $y$ is the received symbol, $x_i$ is the transmitted symbol of user $i$, and $z$ is additive white Gaussian noise with zero mean and variance $N$. User $i$ is also subject to an average power constraint $P_i$. The capacity region of the $M$-user Gaussian MAC can be stated as
\begin{equation}\label{gmac capacity}
\mathcal{C}_{\text{GMAC}}=\{\mathbf{R}|\mathbf{R}(U)\leq \gamma\left(\frac{\mathbf{P}(U)}{N}\right),~\forall~U\subseteq E\},
\end{equation}
where $\gamma(x)=0.5\log_2(1+x)$.

\section{Maximum Decodable Subset}
\begin{figure}
\centering \includegraphics[scale=0.75]{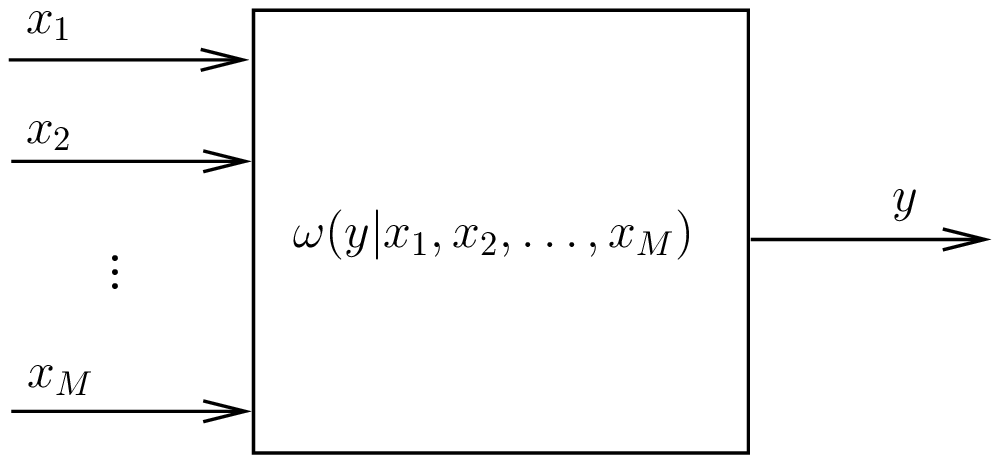}
\caption{Transmitter $i$ uses a random codebook for data transmission at rate $R_i$. Receiver's task is to find the maximum decodable subset of users.}\label{Fig mds}
\end{figure}

In this section, we consider a discrete memoryless channel consisting of $M$ transmitters with input alphabet $\mathscr{X}_i$ for the $i$th transmitter and one receiver with output alphabet $\mathscr{Y}$ where each transmitter uses a single codebook for data transmission. This channel is specified by the transition probability function $\omega(y|x_1,x_2,\ldots,x_M)$ where $x_i\in\mathscr{X}_i$ is the input letter to the channel from the $i$th transmitter and $y\in\mathscr{Y}$ is the output letter received by the receiver, see Figure \ref{Fig mds}. The random codebooks used for data transmission at the rate vector $\mathbf{R}=[R_1,R_2,\ldots,R_M]$ are generated by using the joint probability distribution $p_{X_1}(x_1)p_{X_2}(x_2)\cdots p_{X_M}(x_M)$ for random variables $X_1,X_2,\ldots,X_M$.

The rate vector $\mathbf{R}$ may fall outside of the capacity region of the MAC seen at the receiver side which results in failure to reliably decode all data streams. The receiver's task, however, is to find a decodable subset of transmitters so that their data can be decoded from the received signal. To this end, the receiver partitions the set of transmitters into two disjoint parts and tries to jointly decode the data sent by the transmitters within the first partition, while considering the signals of transmitters in the second partition as noise.

In what follows, we compute the complexity of finding a decodable subset of transmitters by an exhaustive search. Let $E=\{1,2,\ldots,M\}$ denote the set of transmitters' indices. There are $2^{M}$ ways to partition $E$ into two subsets; and to verify that a subset $V$ with cardinality $k$ is decodable, $2^{k}-1$ inequalities must be checked due to (\ref{mac capacity}). Hence, in general, the total number of inequalities to be verified is
\begin{equation}\nonumber
\sum_{k=0}^{M}\binom{M}{k}(2^k-1)= 3^{M}-2^{M},
\end{equation}
which is exponential in the number of users. 

\begin{definition}[Maximum decodable subset]
A set of transmitters is a maximal decodable subset if all transmitters in the subset are jointly decodable by the receiver, and is not a proper subset of any other decodable subset. If the maximal decodable subset is unique, we call it the \emph{maximum decodable subset}.
\end{definition}

\begin{lemma}
For any channel, there is a maximum decodable subset.
\end{lemma}
\begin{proof}
Suppose the receiver is able to decode two subsets of transmitters, namely $U$ and $V$, such that none of them is a subset of the other. $U$ and $V$ are proper subsets of their union $U\cup V$. Besides, their union is decodable by the receiver. This contradicts the fact that both subsets are maximal.
\end{proof}

We first describe some properties of the maximum decodable subset. There are two cases of special interest. The first case occurs when all transmitters are decodable by the receiver, i.e., the maximum decodable subset is the set $E$. In this case, the transmitters' rates must satisfy the inequalities given in (\ref{mac capacity}). In the second case, however, none of the transmitters is decodable by the receiver, i.e., the maximum decodable subset is empty. The following Lemma shows that for the second case the rate vector $\mathbf{R}$ must be in a certain contra-polymatroid.

\begin{lemma}\label{lemma non-decodable}
None of the signals is decodable by the receiver if and only if transmitters' rates satisfy
\begin{equation}\label{contra}
\mathbf{R}(U) >
I(\mathbf{X}_U;Y),~\forall~ U\subseteq E.
\end{equation}
Moreover, the region of the rate vectors satisfying above inequalities forms a contra-polymatroid.
\end{lemma}
\begin{proof}
We first prove that if a rate vector $\mathbf{R}$ satisfies (\ref{contra}), then none of the signals are decodable. To this end, we assume that $V$ is the maximum decodable subset and $V\neq \emptyset$. Since $V$ is a decodable subset, we have the following constraints on the rates of the members of $V$.
\begin{equation}
\mathbf{R}(T)\leq I(\mathbf{X}_T;Y|\mathbf{X}_{V\backslash T}),~\forall~T\subseteq V.
\end{equation}
By substituting $T=V$ in the above equation, we have
\begin{equation}
\mathbf{R}(V)\leq I(\mathbf{X}_V;Y),
\end{equation}
which is a contradiction and this completes the ``if'' part of the proof.

Next, we need to prove that if the inequalities in (\ref{contra}) are not satisfied, there is at least a transmitter which is decodable. Suppose there are some subsets that do not satisfy (\ref{contra}). Assume $W$ has the minimum cardinality among all and satisfies
\begin{equation}\label{gamma-decompose}
\mathbf{R}(W)\leq I(\mathbf{X}_W;Y).
\end{equation}
If $|W|=1$, then the transmitter in $W$ is decodable by considering everything else as noise which is the desired result. Hence, we assume $|W|>1$. If all members of $W$ are jointly decodable, then we have found a decodable subset. Otherwise, there must be a subset of $W$, say $V$, satisfying
\begin{equation}\label{ineq3}
\mathbf{R}(V)>I(\mathbf{X}_V;Y|\mathbf{X}_{W\backslash V}).
\end{equation}
By decomposing the mutual information in (\ref{gamma-decompose}), we obtain
\begin{equation}
\mathbf{R}(W)\leq I(\mathbf{X}_V;Y|\mathbf{X}_{W\backslash V})+I(\mathbf{X}_{W\backslash V};Y).
\end{equation}
From the minimality of $|W|$, we have
\begin{equation}\label{ineq4}
\mathbf{R}(W\backslash V)>I(\mathbf{X}_{W\backslash V};Y).
\end{equation}
By combining the two inequalities (\ref{ineq3}) and (\ref{ineq4}) and considering the fact that $\mathbf{R}(W)=\mathbf{R}(W\backslash V)+\mathbf{R}(V)$, we conclude that
\begin{IEEEeqnarray}{rl}
\mathbf{R}(W)&>I(\mathbf{X}_{W\backslash V};Y)+I(\mathbf{X}_V;Y|\mathbf{X}_{W\backslash V}),\\
&> I(\mathbf{X}_{W};Y),
\end{IEEEeqnarray}
which is a contradiction. This completes the ``only if'' part of the proof.

It is easy to see that the function on the right hand side of (\ref{contra}) is a suppermodular function and monotone, hence the region formed by rates satisfying (\ref{contra}) is a contra-polymatroid.
\end{proof}

In the following theorem, the characterization of the maximum decodable subset is presented.

\begin{theorem}\label{theorem mds}
A subset $S\subseteq E$ is a maximum decodable subset if and only if the transmitters' rates satisfy the following inequalities
\begin{IEEEeqnarray}{rll}
\label{ineq1}\mathbf{R}(V)&\leq I(\mathbf{X}_V;Y|\mathbf{X}_{S\backslash V}),&~\forall~ V\subseteq S,\label{maximum decodable subset relation1} \\
\label{ineq2}\mathbf{R}(U) &>I(\mathbf{X}_U;Y|\mathbf{X}_{S}),&~\forall~ U\subseteq \overline{S}.\label{maximum decodable subset relation2}
\end{IEEEeqnarray}
\end{theorem}
\begin{proof}
Inequality (\ref{ineq1}) corresponds to the capacity region of the MAC for members of $S$ considering members of $\overline{S}$ as noise. Hence, the members of $S$ are decodable iff the inequalities in (\ref{ineq1}) are satisfied. The set $S$ is a maximum decodable subset if no other transmitters in $\overline{S}$ is decodable by the receiver. Now, by applying Lemma \ref{lemma non-decodable} and considering that all members of $S$ are decoded, we conclude that none of the transmitters in $\overline{S}$ is decodable iff the inequalities in (\ref{ineq2}) are satisfied. This completes the proof.
\end{proof}

\begin{figure}
\centering \includegraphics[scale=0.5]{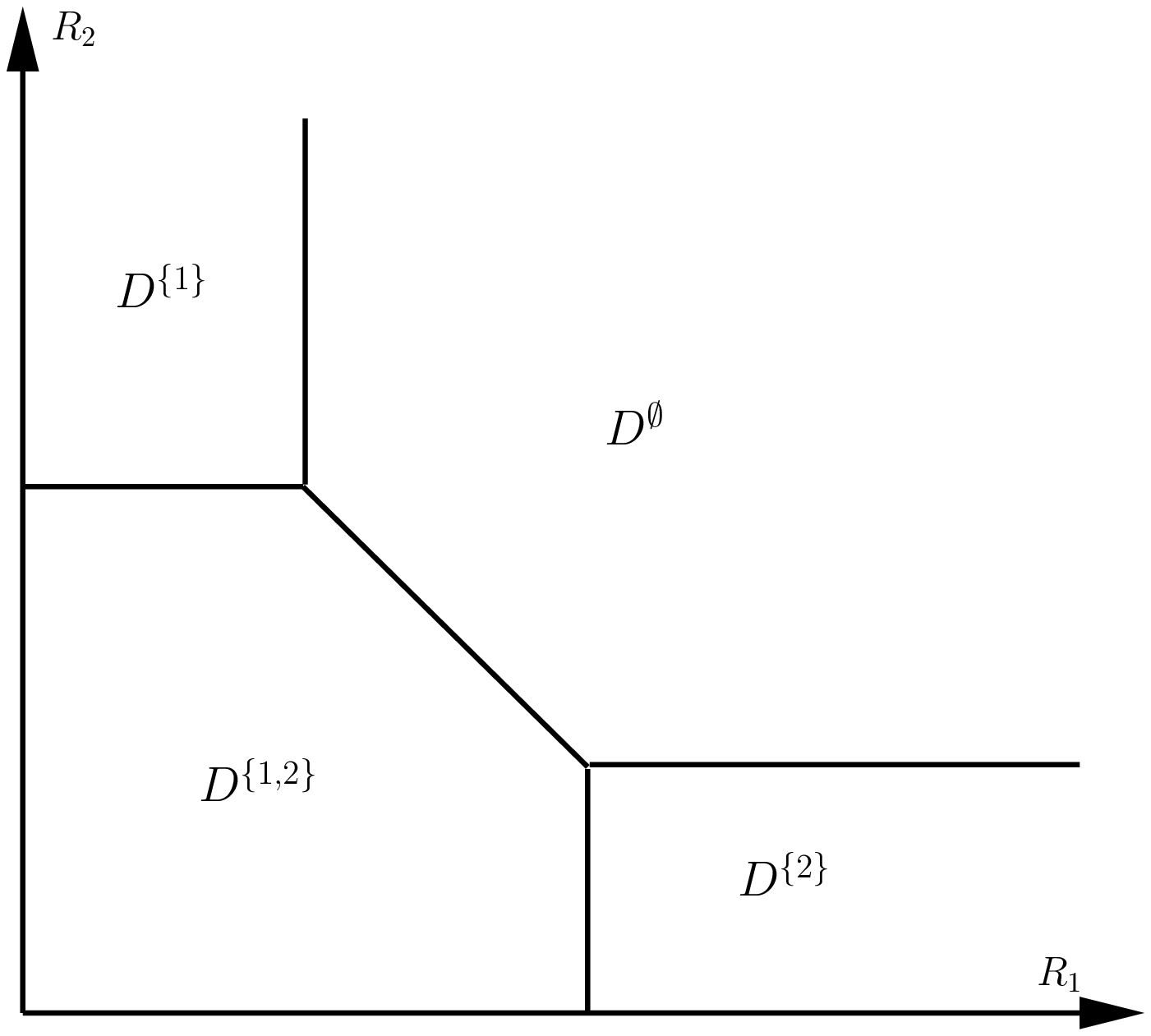}
\caption{Decision regions used for determining the maximum decodable subset for a two-user additive Gaussian Channel. For any rate in $D^{\{1,2\}}$, the receiver can decode both signals. For rates in $D^{\{1\}}$ and $D^{\{2\}}$, the receiver is able to decode transmitters 1 and 2, respectively. Finally, the receiver can decode neither 1 nor 2 for any rate in $D^{\emptyset}$.}\label{two_user_regions}
\end{figure}

For a given maximum decodable subset $S\subseteq E$, we define $D^S$ as
\begin{IEEEeqnarray}{rll}
D^S=\{\mathbf{R}|&\mathbf{R}(T)\leq I(\mathbf{X}_V;Y|\mathbf{X}_{S\backslash V}),&~\forall ~ V\subseteq S,\nonumber\\
&\mathbf{R}(U) >I(\mathbf{X}_U;Y|\mathbf{X}_{S}),&~\forall~ U\subseteq \overline{S}\}.\label{Def.Ds1}
\end{IEEEeqnarray}
$D^S$ is a polyhedron because it is the intersection of finitely many half spaces. By Theorem \ref{theorem mds}, $D^S$ consists of all rate vectors with the same maximum decodable subset $S$. Since for any rate vector there is an associated maximum decodable subset, we have $\cup_{S\subseteq E}D^S=\Re_+^{M}$. This means that $\Re_+^{M}$ is represented as the union of finitely many polyhedral sets. An example for the case of the additive two-user Gaussian channel is given in Figure \ref{two_user_regions}.

The result of this section can be directly extended to continuous channels. The most applicable class of continuous channels is the additive Gaussian channel defined by
\begin{equation}\label{decodable-subset}
y=x_1+x_2+\cdot\cdot\cdot+x_{M}+z,
\end{equation}
where $z$ is an additive Gaussian noise with zero mean and variance $N$. We assume users transmit at rates $\mathbf{R}=[R_1,\ldots,R_{M}]$ using Gaussian codebooks
with average powers $\mathbf{P}=[P_1,\ldots,P_{M}]$. In the following example, we apply the result of Theorem \ref{theorem mds} to a two-user additive Gaussian channel.

\begin{example}
Consider a two-user additive Gaussian channel where the received signal can be written as $y=x_1+x_2+z$. In this case, $E$ has four subsets, namely $S_1=\{1,2\}$, $S_2=\{1\}$, $S_3=\{2\}$, and $S_4=\emptyset$. By applying Theorem \ref{theorem mds}, we obtain the following conditions for the subsets of $E$ to be the maximum decodable subset.
\begin{enumerate}
\item $S_1$ is the maximum decodable subset. In this case, the conditions $R_1\leq \gamma{(\frac{P_1}{N})}$, $R_2\leq \gamma{(\frac{P_2}{N})}$, and $R_1+R_2\leq \gamma{(\frac{P_1+P_2}{N})}$ must be satisfied.
\item $S_2$ is the maximum decodable subset. In this case, the conditions $R_1\leq \gamma\left(\frac{P_1}{P_2+N}\right)$ and $R_2>\gamma\left(\frac{P_2}{N}\right)$ must be satisfied.
\item $S_3$ is the maximum decodable subset. In this case, the conditions $R_2\leq \gamma\left(\frac{P_2}{P_1+N}\right)$ and $R_1>\gamma\left(\frac{P_1}{N}\right)$ must be satisfied.
\item $S_4$ is the maximum decodable subset. In this case, the conditions $R_1> \gamma{(\frac{P_1}{P_2+N})}$, $R_2>\gamma{(\frac{P_2}{P_1+N})}$, and $R_1+R_2>\gamma{(\frac{P_1+P_2}{N})}$ must be satisfied.
\end{enumerate}

The set of conditions described above partitions $\Re_+^2$ into four regions, as illustrated in Figure \ref{two_user_regions}. It can be seen from the figure that $D^{\{1,2\}}$ is a polymatroid corresponding to the capacity region of a two-user MAC and $D^{\emptyset}$ is a contra-polymatroid according to Lemma \ref{lemma non-decodable}.
\end{example}

The above example shows that finding the maximum decodable subset is equivalent to finding the region where the transmitters' rate vector belongs to. Since the number of regions grows exponentially with the number of transmitters, finding a polynomial-time algorithm for solving the problem is desired. To this end, we first define the function $f:2^E\rightarrow\mathbb{R}$ as follows
\begin{equation}\label{sub-min}
f(V)=I(\mathbf{X}_V;Y|\mathbf{X}_{\overline{V}})-\mathbf{R}(V),
\end{equation}
where $V\subseteq E$.

\begin{lemma}
The function $f$ defined in (\ref{sub-min}) is a submodular function.
\end{lemma}

\begin{proof}
The result directly follows from the modularity of $\mathbf{R}$ and the submodularity of mutual information.
\end{proof}

Since there are polynomial-time algorithms for minimizing any submodular functions, c.f., \cite{Schrijver} and \cite{IWATA},  the following optimization problem can be solved efficiently:
\begin{equation}\label{prob1}
f(W)=\min_{V\subseteq E}f(V).
\end{equation}
If the minimum of $f$ in (\ref{prob1}) is zero, then all transmitters are decodable by the receiver due to (\ref{mac capacity}). Otherwise, there is at least a transmitter of the set $E$ which is not decodable. In the following theorem, we prove that indeed all members of the minimizer of $f$ are not decodable, and they need to be considered as noise.
\begin{theorem}\label{theorem noise assumption}
No member of the subset $W$ that minimizes $f$ in (\ref{prob1}) is decodable by the receiver, provided that the minimum in (\ref{prob1}) is not zero and the minimum cardinal minimizer is used. In fact, all members of $W$ must be considered as noise, i.e., if $S$ is the maximum decodable subset then $W\cap S=\varnothing$.
\end{theorem}
\begin{figure}
\centering \includegraphics[scale=0.5]{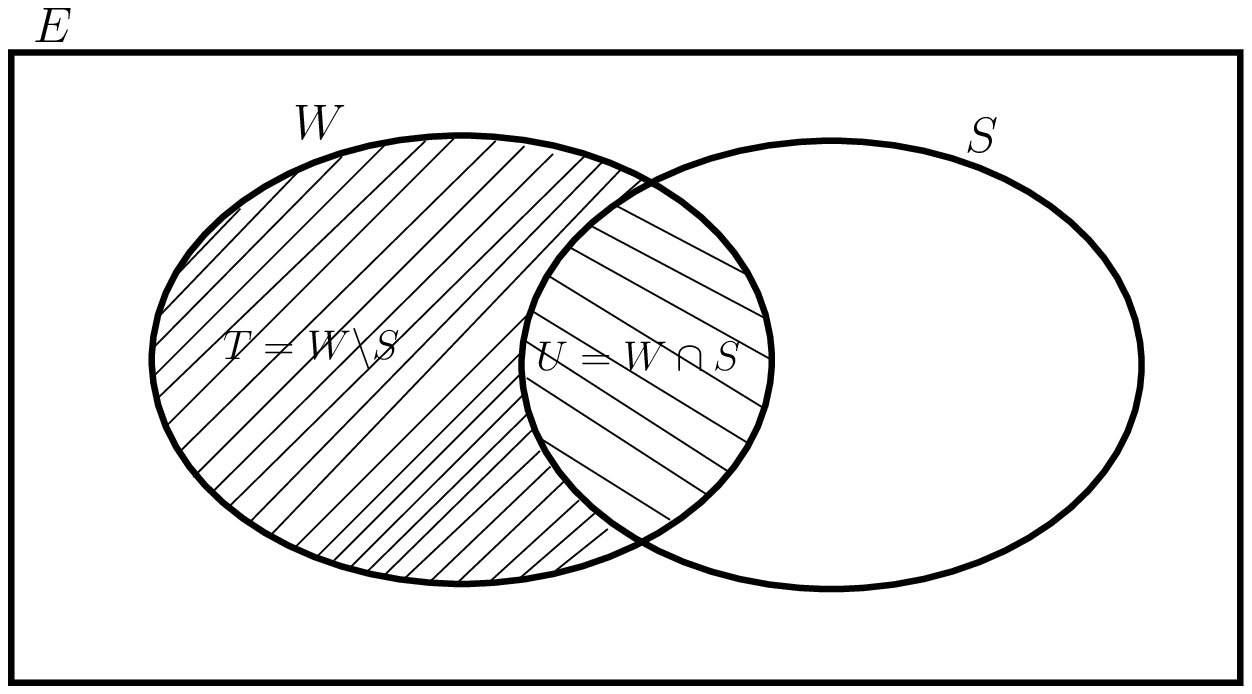}
\caption{$E$ is the ground set. $S$ is the maximum decodable subset. $W$ is the minimizer of $f$ in (\ref{prob1}).}\label{set1}
\end{figure}
\begin{proof}
We first partition the minimizer subset $W$ into two disjoint sets $U$ and $T$ where $U=W\cap S$ and $T=W\backslash S$, see Figure \ref{set1}. We need to show that $U=\emptyset$. Suppose $U$ is nonempty. Hence $|U|\geq 1$ and $|T|<|W|$. Since $U$ is a subset of $S$, from (\ref{maximum decodable subset relation1}), we have
\begin{equation}
\mathbf{R}(U)\leq I(\mathbf{X}_U;Y|\mathbf{X}_{S\backslash U}).
\end{equation}
The inclusion $S\backslash U\subseteq \overline{W}$ and independence of random variables imply $I(\mathbf{X}_U;Y|\mathbf{X}_{S\backslash U})\leq I(\mathbf{X}_U;Y|\mathbf{X}_{\overline{W}})$. Hence,
\begin{equation}\label{positive}
\mathbf{R}(U)\leq I(\mathbf{X}_U;Y|\mathbf{X}_{\overline{W}}).
\end{equation}
From the definition of $f$ in (\ref{sub-min}), we have
\begin{equation}\label{sub min 1}
f(W)=I(\mathbf{X}_W;Y|\mathbf{X}_{\overline{W}})-\mathbf{R}(W).
\end{equation}
From the chain rule and the fact that $T$ and $U$ partition $W$ into two disjoint subsets, we have the following equation
\begin{IEEEeqnarray}{rl}
I(\mathbf{X}_W;Y|\mathbf{X}_{\overline{W}})&=I(\mathbf{X}_T;Y|\mathbf{X}_{\overline{W}},\mathbf{X}_U)
+I(\mathbf{X}_U;Y|\mathbf{X}_{\overline{W}}),\nonumber\\
&=I(\mathbf{X}_T;Y|\mathbf{X}_{\overline{T}})+I(\mathbf{X}_U;Y|\mathbf{X}_{\overline{W}}).\label{temp11}
\end{IEEEeqnarray}
Substituting (\ref{temp11}) into (\ref{sub min 1}) and using $\mathbf{R}(W)=\mathbf{R}(T)+\mathbf{R}(U)$, we obtain
\begin{equation}
f(W)=f(T)+I(\mathbf{X}_U;Y|\mathbf{X}_{\overline{W}})-\mathbf{R}(U).
\end{equation}
Using the inequality (\ref{positive}), we conclude that
\begin{equation}
f(T)\leq f(W).
\end{equation}
If $f(T)< f(W)$, then it contradicts the optimality of $W$, and if $f(T)=f(W)$, then it contradicts the fact that $|W|$ has minimum cardinality among all minimizers. This completes the proof.
\end{proof}

By applying Theorem \ref{theorem noise assumption} and using the well-known submodular function minimization algorithms as a subroutine, c.f. \cite{IWATA} and \cite{Schrijver}, we propose the following polynomial-time algorithm for finding the maximum decodable subset.
\begin{algorithm}[Finding the maximum decodable subset]\label{algorithm1}
~
\begin{enumerate}
\item Set $S=E$.
\item Find $W$ such that
\begin{equation}\nonumber
f(W)=\min_{V\subseteq S}f(V),
\end{equation}
where $f$ is
\begin{equation}
f(V)=I(\mathbf{X}_V;Y|\mathbf{X}_{S\backslash V})-\mathbf{R}(V).
\end{equation}
\item If $W=\varnothing$ STOP. $S$ is the maximal decodable subset. Otherwise, $S\backslash W\longrightarrow S$.
\item If $S=\varnothing$ STOP. No subsect of $E$ is decodable. Otherwise,
GO TO step 2.
\end{enumerate}
\end{algorithm}

\begin{theorem}
Algorithm \ref{algorithm1} converges to the maximum decodable subset in polynomial time.
\end{theorem}
\begin{proof}
Since in each iteration $W$ is a nonempty set (otherwise, the algorithm stops), this algorithm converges at most in $|E|$ iterations. Furthermore, in each iteration, we need to minimize a submodular function which can be done in polynomial time \cite{Schrijver}. Hence, the total running time of the algorithm is polynomial in time.
\end{proof}

\section{An Achievable Rate}
In this section, we propose a method to obtain an achievable rate for the channel $\mathscr{S}$. We also provide a polynomial time algorithm to characterize this achievable rate. A lower bound for the capacity of $\mathscr{S}$ can be obtained by considering interfering users in $E$ as noise and optimizing over all input distributions. Hence, we have
\begin{equation}
\max_{p(x_1)}I(X_1;Y_1)\leq C,
\end{equation}
where $C$ denotes the capacity of $\mathscr{S}$. Now, we assume that regardless of the input distribution, the receiver is able to decode all interfering users considering its own signal as noise. By this assumption, an upper bound on the capacity can be obtained as follows
\begin{equation}
C \leq \max_{p(x_1)}I(X_1;Y_1|\mathbf{X}_{E\backslash 1}).
\end{equation}

Let us assume that the transmitter uses $p_{X_1}(x_1)$ to generate a single random codebook. We need to find the maximum achievable rate $R_1$. If $R_1$ is an achievable rate, then the receiver can successfully decode its intended data. After decoding its own signal, the receiver can search in the set $E\backslash \{1\}$ for the maximum decodable subset $S\subseteq E\backslash \{1\}$. This procedure can be done efficiently using Algorithm \ref{algorithm1}. Let us define $V=E\backslash (S\cup\{1\})$. $V$ is the set of users that receiver treats as noise. From (\ref{maximum decodable subset relation1}), we have
\begin{equation}\label{eq.1}
\mathbf{R}(U)\leq I(\mathbf{X}_U;Y_1|\mathbf{X}_{S\cup\{1\}\backslash U}),~\forall U\subseteq S.
\end{equation}


To find $R_1$, we consider the MAC consisting of user 1 and the users in $S$, while the users in $V$ are considered as noise. From (\ref{mac capacity}), the rate vector $\mathbf{R}$ is achievable if
\begin{equation}\label{eq.2}
\mathbf{R}(U)\leq I(\mathbf{X}_U;Y_1|\mathbf{X}_{S\cup\{1\}\backslash U}),~\forall U\subseteq S\cup \{1\}.
\end{equation}
Since half of the inequalities in (\ref{eq.2}) are satisfied by (\ref{eq.1}) and the only unknown parameter is $R_1$, we can maximize the user's rate based on the following optimization problem:
\begin{IEEEeqnarray}{rl}
R_1(\mathbf{R_{-1}})&=\min_{U\subseteq S} I(X_1,\mathbf{X}_U;Y_1|\mathbf{X}_{S\backslash U})-\mathbf{R}(U).\label{maximum rate}
\end{IEEEeqnarray}
The optimization problem (\ref{maximum rate}) is again a submodular function minimization and can be solved by polynomial-time algorithms.

In the following, we summarize the above procedure.

\begin{algorithm}[finding an achievable rate]\label{algorithm1.1}
~
\begin{enumerate}
\item Given $p(x_1)$, find the maximum decodable subset $S$ among interfering users by using Algorithm \ref{algorithm1} and assuming that the user's data is decoded.
\item Solve the submodular function minimization in (\ref{maximum rate}).
\end{enumerate}
\end{algorithm}

As a by-product of the above algorithm, we can find the subset of interfering users that can be first decoded at the receiver and its effect can be removed.
\begin{proposition}\label{prop1}
If $U$ minimizes (\ref{maximum rate}), then the receiver is capable of decoding all users in $W=S\backslash U$ by considering everything else as noise.
\end{proposition}
\begin{proof}
At the first step, one needs to decode $W$. This requires,
\begin{equation}\label{temp1}
\mathbf{R}(T)\leq I(\mathbf{X}_T;Y_1|\mathbf{X}_{W\backslash T}),~\forall T\subseteq W.
\end{equation}
Suppose there is a subset $T^\star$ that does not satisfy (\ref{temp1}), that is,
\begin{equation}\label{temp2}
\mathbf{R}(T^\star)> I(\mathbf{X}_{T^\star};Y_1|\mathbf{X}_{W\backslash T^\star}).
\end{equation}
Hence,
\begin{IEEEeqnarray}{rl}
\nonumber\tilde{R_1}\stackrel{\bigtriangleup}{=}&I(X_1,\mathbf{X}_{U\cup T^\star};Y_1|\mathbf{X}_{S\backslash ({U\cup T^\star})})-\mathbf{R}({U\cup T^\star})\\
\nonumber\stackrel{(a)}{=}&I(X_1,\mathbf{X}_{U};Y_1|\mathbf{X}_{S\backslash U})-\mathbf{R}(U)\\
\nonumber&+I(\mathbf{X}_{T^\star};Y_1|\mathbf{X}_{W\backslash T^\star})-\mathbf{R}(T^\star)\\
\nonumber\stackrel{(b)}{=}& R_1(\mathbf{R}_{-1})+I(\mathbf{X}_{T^\star};Y_1|\mathbf{X}_{W\backslash T^\star})-\mathbf{R}(T^\star)\\
\label{temp3}\stackrel{(c)}{<} & R_1(\mathbf{R}_{-1}),
\end{IEEEeqnarray}
where (a) follows from the chain rule and the fact that $(S\backslash ({U\cup T^\star}))\cup T^\star=S\backslash U$ and $S\backslash ({U\cup T^\star})=W\backslash T^\star$, (b) follows from the definition of $R_1(\mathbf{R}_{-1})$ and minimality of $U$, and (c) follows form (\ref{temp2}). The last inequality contradicts the fact that $U$ is the solution for the minimization problem in (\ref{maximum rate}). This completes the proof.
\end{proof}

In light of Proposition \ref{prop1}, the set $E$ is decomposable into three disjoint subsets, namely $V$, $U\cup \{1\}$, and $W$. $V$ is the complement of $S\cup \{1\}$, namely the union of the maximum decodable subset $S$ and the intended user. Therefore, the receiver is not able to decode the interfering users in $V$ and considers them as noise. $W$ is the part of $S$ that the receiver can decode by considering everything else as noise. $U\cup \{1\}$ is the subset of users that need to be decoded jointly after removing the effect of $W$.

As indicated in (\ref{maximum rate}), the achievable rate is a function of interfering users' rates. In order to derive some properties of this function, we need the following definition.
\begin{definition}[piecewise linear functions \cite{Rockafellar-wets}]
A function $f:\Re^M\rightarrow\Re$ is piecewise linear if firstly its domain can be represented as the union of finitely many polyhedral sets, and secondly $f$ is ``affine'' within each polyhedral set, i.e., $f(\mathbf{x})=\mathbf{a}^T\mathbf{x}+b$ for some vector $\mathbf{a}$ and scalar $b$.
\end{definition}

In the following theorem, we summarize some properties of $R_1$ as a function of $\mathbf{R}_{-1}$.
\begin{theorem}\label{theorem continuous}
The function $R_1(\mathbf{R}_{-1})$ defined in (\ref{maximum rate}) is 
 piecewise linear. More precisely, $R_1(\mathbf{R}_{-1})$ consists of at most $3^{M-1}$ collection of affine functions.
\end{theorem}
\begin{proof}
Likewise (\ref{Def.Ds1}), let us define the region $D^S$ as
\begin{IEEEeqnarray}{rl}
D^S=\{\mathbf{R}_{-1}|&\mathbf{R}(T)\leq I(\mathbf{X}_T;Y_1|\mathbf{X}_{S\backslash T},X_1),~\forall ~ T\subseteq S,\nonumber\\
&\mathbf{R}(U) >I(\mathbf{X}_U;Y_1|\mathbf{X}_{S},X_1),~\forall~ U\subseteq V\},\label{Def.Ds}
\end{IEEEeqnarray}
where $V=E\backslash (S\cup\{1\})$. Due to (\ref{maximum rate}), the function $R_1(\mathbf{R}_{-1})$ is defined as the pointwise minimum of $2^{|S|}$ affine functions over the polyhedral set $D^S$. As a result, $R_1(\mathbf{R}_{-1})$ is piecewise linear, continuous, and concave over $D^S$, c.f., Theorem 2.49 in \cite{Rockafellar-wets}.

Since $R_1(\mathbf{R}_{-1})$ is a piecewise linear function over each $D^S$ and $\cup_{S\subseteq E\backslash \{1\}}D^S=\Re_+^{M-1}$, it is a piecewise linear function over $\Re_+^{M-1}$. Moreover, each polyhedron $D^S$ is divided into at most $2^{|S|}$ sub-polyhedra in each of which $R_1$ is an affine function. Hence, the total number of components is not more than
\begin{equation}
\sum_{|S|=0}^{M-1}2^{|S|}\binom{M-1}{|S|}=3^{M-1}.
\end{equation}
This completes the proof.
\end{proof}

It is worth noting that, although $R_1$ is a concave function over each $D^S$, it is not a concave function over $\Re_+^{M-1}$.

\begin{figure}
\centering \includegraphics[scale=0.5]{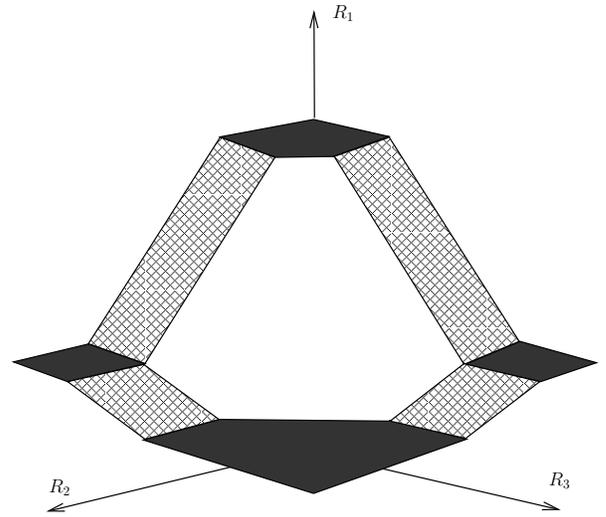}
\caption{The function $R_1(\mathbf{R}_{-1})$ for a channel with two interfering users}\label{piecewise function}
\end{figure}

\begin{example}
Consider an additive channel $y_1=x_1+x_2+x_3+z_1$ where all users use Gaussian codebooks for data transmission. In this case, the maximum decodable subset of interfering users is a subset of $\{2,3\}$. Hence, there are four regions $D^\emptyset$, $D^{\{2\}}$, $D^{\{3\}}$, and $D^{\{2,3\}}$ where $R_1(R_2,R_3)$ is a concave function over each of them. For instance, $R_1(R_2,R_3)=\gamma\left(\frac{P_1}{P_2+P_3+N_1}\right)$ over $D^\emptyset$ and $R_1(R_2,R_3)=\gamma\left(\frac{P_1}{P_2+N_1}\right)-g(R_3)$ over $D^{\{3\}}$ where $g(R_3)$ is either $R_3$ or 0. In Fig. \ref{piecewise function}, an example of the function $R_1(R_2,R_3)$ for this channel is illustrated. As depicted in the figure, $R_1(R_2,R_3)$ is a piecewise linear and continuous function. It also consists of $9$ components, i.e., $3^{M-1}$ for $M=3$.
\end{example}

\begin{example}
In this example, we consider binary adder channel with $M-1$ interfering users. The channel model can be written as $y_1=x_1\oplus x_2\oplus \ldots\oplus x_M$. We further assume that users' codebooks are randomly chosen from Bernouli sequences with $p(0)=p(1)=0.5$. In this case, it is easy to show that
\begin{equation}
R_1(\mathbf{R}_{-1})=\left[1-\mathbf{R}(E\backslash\{1\})\right]^+,
\end{equation}
where $[a]^+=a$ if $a\geq0$ and 0 otherwise. This reflects the fact that the function $R_1(\mathbf{R}_{-1})$ may have less than $3^{M-1}$ components.
\end{example}

%

\section{Channel's Capacity for the Gaussian Case}
In this section, we prove that provided using Gaussian distribution for codebook generation, the achievable rate obtained in the previous section is indeed the capacity of the additive channel with Gaussian noise and $M-1$ Gaussian interfering users.

To show that any rate above $C$ (the output of Algorithm \ref{algorithm1.1} where $p(x_1)$ is Gaussian) is not achievable, we construct a degraded broadcast channel and show that if a rate $R_1>C$ is achievable, then one can communicate reliably outside the capacity region of this channel which is a contradiction. The following lemma assists us in constructing such a degraded channel.

\begin{lemma}\label{broadcasting}
For any set of independent Gaussian codebooks with power vector $\mathbf{P}=[P_1,P_2,\ldots,P_K]$ and rate vector $\mathbf{R}=[R_1,R_2,\ldots,R_K]$, there is a $K$-user Gaussian broadcast channel with the following properties:
\begin{enumerate}
\item The transmitter's total power is $\mathbf{P}(E)$.
\item There are $L$ noise levels: $N_1< N_2<\ldots< N_L$.
\item Users are partitioned into $L$ disjoint subsets, that is, $E=\bigcup_{i=1}^L U_i$. All users in $U_i$ have the same noise level $N_i$, for $i=\{1,2,\ldots,L\}$.
\item The rate vector $\mathbf{R}$ lies on the boundary of the capacity region. $\mathbf{R}$ is achievable using Gaussian codebooks with powers in one to one correspondence with the components of $\mathbf{P}$.
\end{enumerate}
\end{lemma}
\begin{proof}
We aim at building a Gaussian broadcast channel with $x$ as input and $\hat{y}_1,\hat{y}_2,\ldots,\hat{y}_L$ as outputs, where $\hat{y}_i=x+n_i$ and $n_i$ is additive white Gaussian noise with variance $N_i$. To this end, assume there is a $K$-user Gaussian MAC with noise level $N$ and transmit power vector $\mathbf{P}$. Then, $\mathbf{R}$ is achievable if
\begin{equation}\label{Gaussian bc 1}
\mathbf{R}(T)\leq\gamma\left(\frac{\mathbf{P}(T)}{N}\right),~\forall T\subseteq E.
\end{equation}
By monotonicity of $\gamma$, it is always possible to find an $N$ such that the rate vector $\mathbf{R}$ is achievable. Indeed, $\mathbf{R}$ is achievable for any $N\in [0,N_1]$, where $N_1$ corresponds to the case that for any noise above $N_1$ at least one of the inequalities in (\ref{Gaussian bc 1}) turns to equality. Let $U_1$ denote the set of users for which the corresponding inequality in (\ref{Gaussian bc 1}) turns to equality with noise level $N_1$ (in case of having more than one equality we choose the maximum cardinal subset), i.e.,
\begin{equation}\label{Gaussian bc 2}
\mathbf{R}(U_1)=\gamma\left(\frac{\mathbf{P}(U_1)}{N_1}\right).
\end{equation}
By plugging in (\ref{Gaussian bc 1}), we have
\begin{equation}\label{Gaussian bc 3}
\mathbf{R}(T)<\gamma\left(\frac{\mathbf{P}(T)}{N_1+\mathbf{P}(U_1)}\right),~\forall T\subseteq E\backslash U_1.
\end{equation}
Now, we correspond users in $U_1$ to the output of the Gaussian channel $\hat{y}_1=x+n_1$, where $n_1$ is additive Gaussian noise with variance $N_1$.

We  can apply the same procedure to (\ref{Gaussian bc 3}), i.e., we increase $N_1$ until one of the inequalities turns to equality. Let $N_2$ denote the maximum noise level satisfying (\ref{Gaussian bc 3}) with equality. Clearly, $N_1<N_2$. If $U_2$ denotes the set of users satisfying (\ref{Gaussian bc 3}) with equality, then we have
\begin{equation}\label{Gaussian bc 4}
\mathbf{R}(U_2)=\gamma\left(\frac{\mathbf{P}(U_2)}{N_2+\mathbf{P}(U_1)}\right).
\end{equation}
By plugging in (\ref{Gaussian bc 3}), we obtain
\begin{equation}\label{Gaussian bc 5}
\mathbf{R}(T)<\gamma\left(\frac{\mathbf{P}(T)}{N_1+\mathbf{P}(U_1)+\mathbf{P}(U_2)}\right),~\forall T\subseteq E\backslash U_1\cup U_2.
\end{equation}
Now, we correspond users in $U_2$ to the output of the Gaussian channel $\hat{y}_2=x+n_2$, where $n_2$ is additive Gaussian noise with variance $N_2$.

By repeating the above procedure, we can construct a set of channels with noise levels $N_1<N_2<\ldots<N_L$ and associate set of users $U_1,U_2,\ldots,U_L$ with $E=\bigcup_{j=1}^{L}U_j$ such that
\begin{IEEEeqnarray}{rl}\label{Gaussian bc 6}
\mathbf{R}(U_i)&=\gamma\left(\frac{\mathbf{P}(U_i)}{N_i+\mathbf{P}(\bigcup_{j=1}^{i-1}U_j)}\right),\\
\mathbf{R}(T)&\leq\gamma\left(\frac{\mathbf{P}(T)}{N_i+\mathbf{P}(\bigcup_{j=1}^{i-1}U_j)}\right),~\forall T\subseteq U_i,\label{Gaussian bc 7}\\
\mathbf{R}(T)&<\gamma\left(\frac{\mathbf{P}(T)}{N_i+\mathbf{P}(\bigcup_{j=1}^{i}U_j)}\right),~\forall T\subseteq \bigcup_{j=i+1}^{L}U_j.\label{Gaussian bc 8}
\end{IEEEeqnarray}

Now, assume that the transmitter with total power $\mathbf{P}(E)$ uses $K$-level Gaussian codebooks for data broadcasting. The transmitted signal can be written as $\mathbf{x}=\mathbf{x}_1+\mathbf{x}_2+\ldots+\mathbf{x}_K$, where $\mathbf{x}_l$ is a Gaussian codeword with power $P_l$ and rate $R_l$ and contains information for $l$'th user. The received signal at noise level $N_i$ can be written as $\hat{\mathbf{y}}_i=\mathbf{x}+\mathbf{n}_i$. The set of inequalities in (\ref{Gaussian bc 8}) implies that all users at noise level $N_i$ can decode data streams of users in $\bigcup_{j=i+1}^L U_j$ considering everything else as noise. By removing the effect of users in $\bigcup_{j=i+1}^L U_j$ from the received signal, the set of inequalities in (\ref{Gaussian bc 7}) implies that all users in $U_i$ can decode their own signal considering users in $\bigcup_{j=1}^{i-1} U_j$ as noise. In other words, all users at the same level of noise can decode their signals by first decoding the users at upper levels and removing their effect and considering users at lower levels as Gaussian noise. Hence, we obtain a Gaussian broadcast channel in which the rate vector $\mathbf{R}$ is achievable and Gaussian codebooks are constructed according to the power vector $\mathbf{P}$. It remains to show that $\mathbf{R}$ is on the boundary of the capacity region. The capacity region of the Gaussian broadcast channel is fully characterized and there is an explicit expression for boundary points, c.f. \cite{Cover-thomas}. The equalities in (\ref{Gaussian bc 6}) guarantee that the rate vector $\mathbf{R}$ lies on the boundary of the capacity region. This completes the proof.
\end{proof}

\begin{theorem}\label{theorem capacity}
The rate $C$, the output of Algorithm \ref{algorithm1.1}, is the capacity of the channel described in (\ref{single-receiver}).
\end{theorem}
\begin{proof}
We rewrite the achievable rate given in Algorithm \ref{algorithm1.1} by using the Gaussian distribution as codebook generator. As discussed earlier, the set of users can be partitioned into three subsets $V$, $U\cup \{1\}$, and $W$.

$W$ is the subset of interfering users that the receiver can decode considering everything else as noise. Since the Gaussian noise is the worst noise for additive channels, c.f. \cite{Ihara:non-Gaussian} and \cite{Diggavi:worst-noise}, and $W$ is decodable when other users are considered as Gaussian noise, $W$ is decodable for any arbitrary distribution for intended user. As a result, interfering users in $W$ can be completely eliminated regardless of the input codebook.

$V$ is the complement of the maximum decodable subset and must be considered as noise. From (\ref{maximum decodable subset relation2}), we have
\begin{equation}\label{temp4}
\mathbf{R}(T)>\gamma\left(\frac{\mathbf{P}(T)}{N+\mathbf{P}(V\backslash T)}\right),~\forall T\subseteq V.
\end{equation}

$U$ is the solution to the minimization problem in (\ref{maximum rate}). Hence, we have
\begin{IEEEeqnarray}{rl}\label{capacity1}
C+\mathbf{R}(U)&=\gamma\left(\frac{P_1+\mathbf{P}(U)}{N+\mathbf{P}(V)}\right).
\end{IEEEeqnarray}

We apply Lemma \ref{broadcasting} to the set of users in $V$ with associated power vector $\mathbf{P}(V)$ and rate vector $\mathbf{R}(V)$. Let $N_1<N_2<\ldots<N_L$ denote the noise levels and $U_1,U_2,\ldots,U_L$ denote the collection of subsets of users associated to each level of noise for the Gaussian broadcast channel with the properties given in Lemma \ref{broadcasting}. We claim that $N_L<N$. To verify this, we substitute $U_L$ into (\ref{Gaussian bc 6}) and (\ref{temp4}). Hence, we obtain
\begin{equation}
\gamma\left(\frac{\mathbf{P}(U_L)}{N_L+\mathbf{P}(V\backslash U_L)}\right)>\gamma\left(\frac{\mathbf{P}(U_L)}{N+\mathbf{P}(V\backslash U_L)}\right)
\end{equation}
which results in $N_L<N$.

Next, we add $U_{L+1}=U\cup\{1\}$ as a set of new users to the Gaussian broadcast channel with noise level $N_{L+1}=N$ and increase the transmitter's total power by $P_1+\mathbf{P}(U)$. It is easy to verify that the conditions in (\ref{Gaussian bc 6}), (\ref{Gaussian bc 7}), and (\ref{Gaussian bc 8}) are still satisfied with new broadcast channel. Consequently, the rate vector lies on the boundary of the capacity region. Therefore, reliable data transmission at any rate above $C$ results in reliable data transmission outside the capacity region which is a contradiction. This completes the proof.
\end{proof}

\section{Applications for the $M$-user Gaussian IC}

In this section, we apply the proposed algorithms to the $M$-user Gaussian IC modeled by
\begin{equation}
y_i=\sum_{i=1}^Mh_{ij}x_j+z_i,
\end{equation}
where $x_j$ is the transmitted symbol of user $j$ and $h_{ij}$ denotes the link's gain between the $j$'th transmitter and the $i$'th receiver. $z_i$ is additive white Gaussian noise with zero mean and variance $N_i$. User $i$ is also subject to an average power constraint $P_i$. The capacity region of this channel is denoted by $\mathscr{C}_{GIC}$.

It is more convenient to write the system model in matrix form as
\begin{equation}\label{matrix model}
\mathbf{y}=H\mathbf{x}+\mathbf{z},
\end{equation}
where $\mathbf{y}=[y_1,y_2,\ldots,y_M]^T$ and $\mathbf{x}=[x_1,x_2,\ldots,x_M]^T$ denote the output and input vectors, respectively. $H=[h_{ij}]$ is the matrix of links' gains, and $\mathbf{z}=[z_1,z_2,\ldots,z_M]^T$ is the Gaussian noise vector which has a diagonal covariance matrix. By scaling transformations, it is possible to write the channel model (\ref{matrix model}) in standard form where the noise variances and diagonal elements of $H$ are one \cite{IC:Carleial}.

Let us assume each transmitter is allowed to transmit data by using a single Gaussian codebook and each receiver is allowed to decode any subset of interfering users. Let $\Psi$ denote the set of decoding strategies. By a decoding strategy $\psi=\{S_1,S_2,\ldots,S_M\}\in\Psi$, we mean that the receiver $i$ tries to decode all users' data in $S_i$. Clearly, $S_i\subseteq E$ and $i\in S_i$. Since there are $2^{M-1}$ possible choices for each $S_i$, we have $2^{M(M-1)}$ possible strategies in total. Hence, $|\Psi|=2^{M(M-1)}$.

Given a strategy, a rate vector $\mathbf{R}$ is achievable with respect to that strategy if every receiver can reliably decode its associated users. Therefore, an achievable rate region $\mathscr{C}_{\psi}$ can be defined as a set of all rate vectors that are achievable with respect to the strategy $\psi$. Let $\mathscr{C}_o=\bigcup_{\psi\in\Psi}\mathscr{C}_{\psi}$. Clearly, $\mathscr{C}_o\subseteq \mathscr{C}_{GIC}$ and it can be shown that $\mathscr{C}_o$ is not convex in general.

\subsection{Some Extreme Points of $\mathscr{C}_o$}
Given an ordering $\pi$ of users, we aim at maximizing users' rates in accordance with $\pi$. In general, there are $M!$ orderings of users which result in $M!$ not necessarily distinct achievable rates in the capacity region. Due to the polymatroidal property of the capacity region of the Gaussian MAC, every permutation leads to a distinct achievable rate vector; whereas, $\mathscr{C}_o$ is not a polymatroid and hence there may be some permutations that lead to the same achievable rate vector. Without loss of generality, we may assume the order is the same as that of users' indices, i.e., permutation matrix is identity.

Setting the first user's rate to its maximum value $R_1=\gamma\left(P_1\right)$ imposes some constraints on the other user's rates as they must be decoded by the first receiver. The reason is that $R_1$ is achievable if the first receiver can decode all the interfering users by considering its own signal as noise and eliminating their effects from the received signal.

Maximization of the second user's rate is more delicate, since its transmission should not affect the first user's data rate. However, we have the choice of lowering other users' rates as much as needed. Hence, we assume users in the set $\{3,4,\ldots,M\}$ are decoded at the first and second receivers by considering everything else as noise and their effects are removed. $R_2$ must be chosen such that both receivers can decode it. The maximum decodable subset at the first receiver is $\{1\}$ by the assumption. For the second user, we can find the maximum decodable subset of interfering users which in this case is either $\emptyset$ or $\{1\}$. Now, we can run Algorithm \ref{algorithm1.1} at both receivers to find an achievable rate for each receiver. Clearly, the minimum of the two achievable rates are achievable and we set $R_2$ to this value. Besides, we obtain the strategy $\psi^{(2)}=\{S_1^{(2)},S_2^{(2)}\}$ in which $R_1$ and $R_2$ are achievable, where $S_1^{(2)},S_2^{(2)}\subseteq E^{(2)}$ and $E^{(i)}=\{1,2,\ldots,i\}$.

To maximize the rate of user $i$, we proceed as follows. We treat users above index $i$ as they do not exist, i.e., we put constraints on their rates in such a way that all the receivers with indices in $E^{(i)}$ can decode them first and remove their effects. From maximization of users' rates in the previous steps, we have $\mathbf{R}_{E^{(i-1)}}$ and its corresponding achievable strategy $\psi^{(i-1)}=\{S_1^{(i-1)},\ldots,S_{i-1}^{(i-1)}\}$, where $S_j^{(i-1)}\subseteq E^{(i-1)},~\forall j\in E^{(i-1)}$. $R_i$ must be chosen such that all receivers in $E^{(i)}$ can decode it. The maximum decodable subset of interfering users is given by $\psi^{(i-1)}$ for all receivers in $E^{(i-1)}$. We can also find the maximum decodable subset of interfering users at receiver $i$ by running Algorithm \ref{algorithm1}. Let $S_i^{(i-1)}$ denote this subset. From (\ref{maximum rate}), $R_i$ is achievable at the receiver $j\in E^{(i)}$, if it is less than $R_{ij}$ which is defined as
\begin{equation}
R_{ij}=\min_{U\subseteq S_j^{(i-1)}}\gamma\left(\frac{h_{ji}^2P_i+\sum_{k\in U}h_{jk}^2P_k}{1+\sum_{k\in E^{(i-1)}\backslash S_j^{(i-1)} }h_{jk}^2P_k}\right)-\mathbf{R}(U).\label{rateij}
\end{equation}
Hence, $R_i$ can be chosen as the minimum of all $R_{ij}$s. For the next step, we need a new achievable strategy. It is easy to see that $\psi^{(i)}=\{S_1^{(i-1)}\cup \{i\},\ldots,S_{i-1}^{(i-1)}\cup \{i\},S_{i}^{(i-1)}\cup \{i\}\}$ is the proper strategy at step $i$. Now, we can iterate until the last user.

\begin{algorithm}[successive maximization of users' rates]\label{algorithm2}
~
\begin{enumerate}
\item Set $R_1=\gamma\left(P_1\right)$ and $S_1^{(1)}=\{1\}$.
\item For $i=2:M$, do:
\begin{enumerate}
\item Find the maximum decodable subset of interfering users $S_i^{(i-1)}$ in the set $E^{(i-1)}$ for receiver $i$ assuming that users in the set $E\backslash E^{(i-1)}$ are decoded and their effects are removed.
\item Solve the following optimization problem
\begin{equation}\label{rate_order_max}
R_i=\min_{j\in E^{(i)}} R_{ij},
\end{equation}
where $R_{ij}$ is defined in (\ref{rateij}).
\item $S_j^{(i)}=S_j^{(i-1)}\cup \{i\}$, for all $j\in E^{(i)}$.
\end{enumerate}
\end{enumerate}
\end{algorithm}

For the sake of completeness, in the following theorem, we state that the above algorithm finishes in polynomial time.
\begin{theorem}
Algorithm \ref{algorithm2} converges to an extreme point of $\mathscr{C}_o$ in polynomial time.
\end{theorem}
\begin{proof}
At the $i$'th iteration, we need to solve $i$ submodular function minimizations. Hence, in total, a submodular function minimization subroutine is invoked for $M(M+1)/2$ times. Moreover, at each step,  we need to find the maximum decodable subset which can be accomplished in polynomial time based on Theorem \ref{theorem mds}. Hence, Algorithm \ref{algorithm2} converges to an extreme point of $\mathscr{C}_o$ in polynomial time.
\end{proof}

It is worth noting that for the two-user Gaussian IC in the case of strong and very strong interference \cite{IC:SASON}, the output of Algorithm \ref{algorithm2} is a point on the boundary of the capacity region. In the case of weak interference, however, the output of Algorithm \ref{algorithm2} coincides with Costa's result in \cite{IC:COSTA}. Unfortunately, the optimality of the result claimed by Costa has not been proved yet \cite{IC:SASON}. As a result, proving the optimality of extreme points obtained from Algorithm \ref{algorithm2} has at least the same level of difficulty as that of the two-user case.

\subsection{Generalized One-sided Gaussian IC}
Parallel to the definition of the one-sided Gaussian IC \cite{IC:COSTA}, we define the generalized one-sided Gaussian IC as one in which the channel matrix $H$ can be represented as a triangular matrix by row permutations.  For the sake of simplicity, we always assume that $H$ is lower triangular. Hence, the first user incurs no interference from other users, i.e., $y_1=x_1+z_1$, the second user incurs interference only form the first user, i.e., $y_2=h_{21}x_1+x_2+z_2$, and in general, user $i$ incurs interference from preceding users, i.e., $y_i=h_{i1}x_1+\ldots+h_{i(i-1)}x_{i-1}+x_i+z_i$.

From optimality of the proposed achievable rate region for the Gaussian case, we can prove the following theorem.

\begin{theorem}
For the generalized one-sided Gaussian IC, successive maximization of users' rates gives an achievable rate vector which lies on the boundary of the capacity region.
\end{theorem}

\begin{proof}
We can prove this theorem by induction on the number of users. For a single user, it is trivial that maximizing user's rate result in transmission at the capacity of the channel. Now, we assume that for $m-1$ users, successive maximization of users' rates gives us a point $\mathbf{R}_{E^{(m-1)}}$ on the capacity region. Due to Theorem \ref{theorem capacity}, user $m$, which does not interfere with other users and only receives interference from other users, transmits at a rate $R_m$ which is the capacity of its channel assuming other users use Gaussian codebooks for data transmission. Since no further increasing of any rate is possible, the rate vector $\mathbf{R}=[\mathbf{R}_{E^{(m-1)}},R_m]$ lies on the boundary of the capacity region of the $m$-user Gaussian IC.
\end{proof}

The capacity region of strong and very strong two-user Gaussian ICs is known and corresponds to the capacity of the corresponding compound MAC where both receivers decode both users' messages \cite{STRONG:SATO} \cite{Very_Storng:Carleial}. Therefore, for the $M$-user case, it is interesting to find similar situations where the capacity is achievable when all receivers decode all messages sent by all transmitters. However, by a counter example, it is easy to show that having the condition $h_{ij}^2\geq 1,~\forall i,j\in E$, is not sufficient to establish similar results. To find similar situations, we define the strong generalized one-sided Gaussian IC as the channel with triangular channel matrix $H$ with the property that $h_{ik}^2\geq h_{jk}^2$ whenever $i\geq j$. In the following theorem, we prove that the capacity region of the strong generalized one-sided Gaussian IC can be fully characterized.

\begin{theorem}
The capacity region of the strong generalized Z Gaussian IC is $\bigcap_{i\in E} \mathcal{C}_{MAC}(i)$, where $\mathcal{C}_{MAC}(i)$ denotes the capacity region of the MAC seen at the $i$th receiver.
\end{theorem}
\begin{proof}
This theorem can be also proved by induction on the number of users. For a single user, it is trivial. We assume that for a channel with $m-1$ users and a triangular channel matrix, the capacity region is $\bigcap_{i\in E\backslash \{m\}} \mathcal{C}_{MAC}(i)$. Now, we add a new user which does not interfere with other users and only receives interference from all other users. Let $\mathcal{C}_{GIC}$ denote the capacity region of $M$-user Gaussian IC. It suffices to show that for any rate vector $\mathbf{R}=[\mathbf{R}_{-m},R_m]$ in $\mathcal{C}_{GIC}(m)$, receiver $m$ is able to decode all users' messages. The idea that we use here is similar to the idea of Han and Kobayashi for proving the capacity region of strong and very strong two-user Gaussian ICs \cite{IC:HK}. Since $\mathbf{R}_{-m}$ is achievable and there is no interference from user $m$, we have $\mathbf{R}_{-m}\in \bigcap_{i\in E\backslash \{m\}} \mathcal{C}_{MAC}(i)$. In particular, $\mathbf{R}_{-m}\in \mathcal{C}_{MAC}(m-1)$. Hence, receiver $m-1$ which has $y_{m-1}=h_{(m-1)1}x_1+\cdots+h_{(m-1)(m-2)}x_{m-2}+x_{m-1}+z_{m-1}$ as the received signal can jointly decode all users in the set $E\backslash \{m\}$.
Since $R_m$ is decodable by the $m$th receiver, it can be removed from the received signal $y_{m}=h_{m1}x_1+\cdots+h_{M(m-1)}x_{m-1}+x_{m}+z_{m}$. Now, receiver $m$ can try to decode other users' data from $\tilde{y}_{m}=h_{m1}x_1+\cdots+h_{m(m-1)}x_{m-1}+z_{m}$. Let $\mathcal{\tilde{C}}_{MAC}(m-1)$ denote the capacity of this MAC. By hypothesis, $h_{ik}^2\geq h_{jk}^2$ whenever $i\geq j$. Therefore, $\mathcal{\tilde{C}}_{MAC}(m-1)\subseteq \mathcal{{C}}_{MAC}(m-1)$. Hence, receiver $m$ is able to decode the rate vector $\mathbf{R}_{-m}$. This completes the proof.
\end{proof}

\section{Conclusion}
We investigated data transmission over a channel with $M-1$ interfering users.
By establishing certain properties of the maximum decodable subset, we proposed a polynomial time algorithm that separates the interfering users into two disjoint parts: the users that the receiver is able to jointly decode their messages and its complement. We introduced an optimization problem that gives an achievable rate for this channel. We proposed a polynomial time algorithm for solving this optimization problem. We also established the capacity of the additive Gaussian channel with Gaussian interfering users and showed that the Gaussian distribution is optimal and the proposed achievable rate is the capacity of this channel.

As an application of this method, we investigated data transmission for the case of $M$-user interference channel when transmitters use single codebooks for data transmission, and receivers are allowed to decode other users' messages. We then introduced an achievable rate region $\mathscr{C}_o$. We obtained some extreme points of $\mathscr{C}_o$ by successive maximization of users' rates. We also characterized one point on the capacity of the generalized one-sided Gaussian IC. Finally, we obtained the capacity region of the strong generalized one-sided Gaussian IC.

\bibliographystyle{IEEEtran}
\bibliography{to_dedoce_or_not}

\end{document}